\documentclass[12pt]{article}

\usepackage{graphicx}
\usepackage{float}
\usepackage{amsmath}
\usepackage{amssymb}
\usepackage{authblk}
\usepackage[labelsep=period,labelfont={small,bf},font={small}]{caption}

\begin{document}

\title{Comprehensive evaluation of the linear stability of Alfv\'en eigenmodes driven by alpha particles in an ITER baseline scenario}

\author[1]{A.C.A. Figueiredo}
\author[1]{P. Rodrigues} 
\author[1]{D. Borba}
\author[1]{R. Coelho}
\author[1]{L. Fazendeiro}
\author[1]{J. Ferreira}
\author[1,2]{N.F. Loureiro}
\author[1]{F. Nabais}
\author[3]{S.D. Pinches}
\author[3]{A.R. Polevoi}
\author[4]{S.E. Sharapov}

\affil[1]{Instituto de Plasmas e Fus\~{a}o Nuclear, Instituto Superior T\'{e}cnico, Universidade de Lisboa, 1049-001 Lisboa, Portugal}
\affil[2]{Plasma Science and Fusion Center, Massachusetts Institute of Technology, Cambridge MA 02139, USA}
\affil[3]{ITER Organization, Route de Vinon-sur-Verdon, CS 90 046, 13067 St Paul-lez-Durance Cedex, France}
\affil[4]{Culham Centre for Fusion Energy, Culham Science Centre, Abingdon, OX14 3DB, UK}
\affil[ ]{}
\affil[ ]{E-mail: antonio@ipfn.tecnico.ulisboa.pt}

\date{}

\maketitle

\begin{abstract}
The linear stability of Alfv\'en eigenmodes in the presence of fusion-born alpha particles is thoroughly assessed for two variants of an ITER baseline scenario, which differ significantly in their core and pedestal temperatures.
A systematic approach is used that considers all possible eigenmodes for a given magnetic equilibrium and determines their growth rates due to alpha-particle drive and Landau damping on fuel ions, helium ashes and electrons.
This extensive stability study is efficiently conducted through the use of a specialized workflow that profits from the performance of the hybrid MHD drift-kinetic code \mbox{CASTOR-K} (Borba D. and Kerner W. 1999 J. Comput. Phys. {\bf 153} 101; Nabais F. {\it et al} 2015 Plasma Sci. Technol. {\bf 17} 89), which can rapidly evaluate the linear growth rate of an eigenmode.
It is found that the fastest growing instabilities in the aforementioned ITER scenario are core-localized, low-shear toroidal Alfv\'en eigenmodes.
The largest growth-rates occur in the scenario variant with higher core temperatures, which has the highest alpha-particle density and density gradient, for eigenmodes with toroidal mode numbers $n\approx30$.
Although these eigenmodes suffer significant radiative damping, which is also evaluated, their growth rates remain larger than those of the most unstable eigenmodes found in the variant of the ITER baseline scenario with lower core temperatures, which have $n\approx15$ and are not affected by radiative damping.
\end{abstract}

\noindent{\it Keywords}: Alfv\'en eigenmodes, alpha particles, linear stability, burning plasmas, ITER

\section{Introduction}

In ITER, the performance of burning plasmas will depend on the population of alpha particles being well confined within the plasma core, as the heating of the deuterium-tritium (DT) plasma will then rely mainly on the energy of these suprathermal particles that are produced by core fusion reactions.
A phenomenon that can potentially hinder the successful operation of ITER is therefore the destabilization of Alfv\'en eigenmodes (AEs) by alpha particles \cite{Fu1989}, whereby an increased radial transport of the latter could degrade the conditions necessary to sustain the fusion process and lead to power fluxes that exceed the design values of the ITER plasma facing components \cite{Fasoli2007, Kurki-Suonio2009, Sharapov2013}.

While ITER scenario development is underway and in the absence of experimental results for guidance, a comprehensive modelling approach is mandatory to forecast the stability of Alfv\'enic activity in ITER plasmas.
In this article the stability of AEs is systematically addressed for the 15 MA ELMy H-mode ITER baseline scenario \cite{Polevoi2002, Pinches2015, Lauber2015, Rodrigues2015} making use of a recently introduced framework \cite{Rodrigues2015} that is based on the hybrid MHD drift-kinetic code \mbox{CASTOR-K} \cite{Borba1999, Nabais2015}, which is the key element in our suite of numerical codes.
Growth rates are computed systematically for all possible eigenmodes taking into account the alpha-particle drive, the Landau damping due to the DT ions, the thermalized helium (He) ions and the electrons, and the interaction with the Alfv\'en continuum.
Radiative damping is calculated a posteriori for the most relevant modes.

The remainder of this article is organized in three main sections.
The ITER baseline scenario, our modelling workflow and the numerical codes on which it relies are described in section \ref{section_scenario_and_workflow}. 
section \ref{section_results} is dedicated to a detailed discussion of the results.
Finally, a summary and conclusions are provided in section \ref{section_conclusions}.

\section{Scenario and workflow}
\label{section_scenario_and_workflow}

\subsection{ITER baseline scenario}
\label{section_scenario}

\begin{figure*}
\begin{center}
\includegraphics[width=1.0\textwidth]{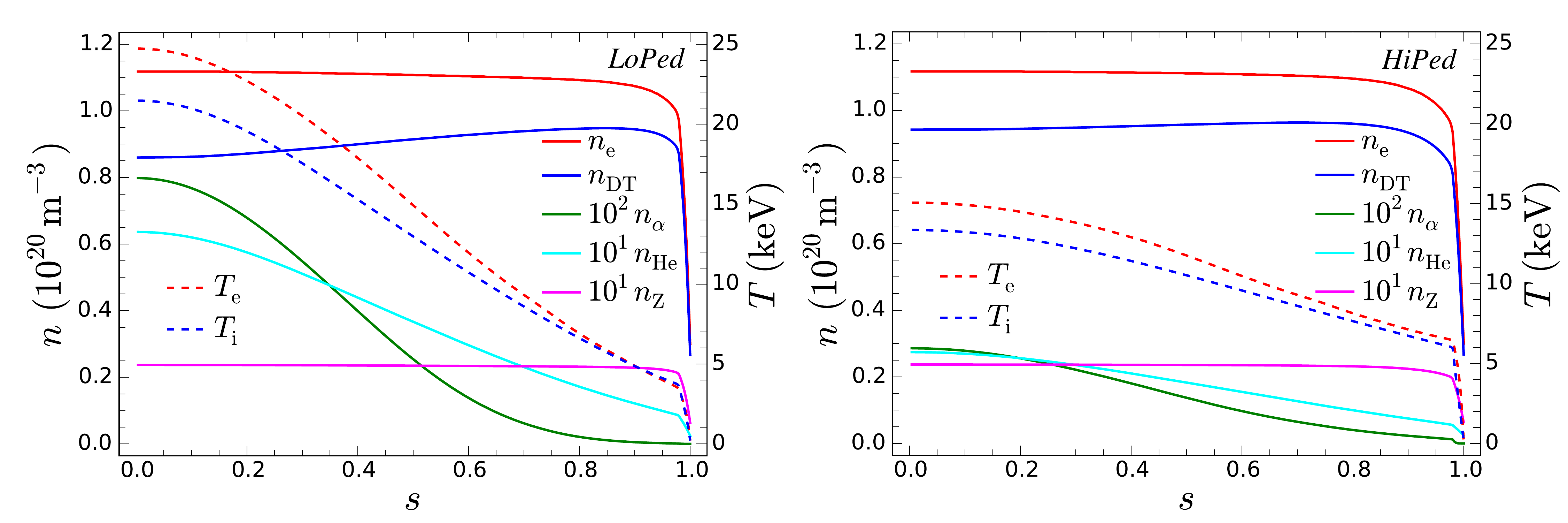}
\caption{Kinetic profiles from ASTRA simulations \cite{Polevoi2002,Pereverzev2008} for two variants of the 15 MA ITER baseline scenario. Compared with the {\it LoPed} scenario variant (left), {\it HiPed} (right) is characterised by higher pedestal temperatures and lower temperatures at the plasma core (dashed lines), which results in lower densities of the main fusion products, namely alpha particles and He ashes.}
\label{figure_scenario}
\end{center}
\end{figure*}
The two variants of the ITER scenario that are analysed here have been produced by ASTRA transport simulations \cite{Polevoi2002,Pereverzev2008}.
In both variants the plasma current is $I_p$ = 15 MA, the toroidal magnetic field is $B_0$ = 5.3 T, and the plasma major and minor radii are $R_0$ = 6.2 m and $a$ = 2 m, respectively.
The position of the magnetic axis is $R_\mathrm{m}$ = 6.4 m.

Figure \ref{figure_scenario} shows density and temperature profiles versus radial coordinate $s=\sqrt{\Psi/\Psi_\mathrm{b}}$, where the poloidal magnetic flux $\Psi$ is normalized to its boundary value $\Psi_\mathrm{b}$.
It can be seen that both scenario variants have approximately the same electron density $n_\mathrm{e}$ and impurity content $n_\mathrm{Z}$, which is essentially due to beryllium coming from the first wall.
The main difference between the two variants is that in the one on the left of figure \ref{figure_scenario}, hereafter referred to as {\it LoPed}, electron ($T_\mathrm{e}$) and ion ($T_\mathrm{i}$) temperatures are much lower at the pedestal and much higher at the core than in the {\it HiPed} variant on the right.
In fact, the {\it LoPed} and {\it HiPed} scenarios cover a 3 keV range of pedestal-top temperatures around the expected value for ITER, which is approximately 5 keV on the basis of edge MHD stability \cite{Huijsmans2013}.
Naturally, the higher (by roughly 10 keV) core temperatures in {\it LoPed} go together with a higher density of alpha particles $n_\alpha$ and of helium ashes (thermalized alpha particles) $n_\mathrm{He}$ in {\it LoPed} than in {\it HiPed}.
This in turn is reflected on the fuel density $n_\mathrm{DT} = n_\mathrm{D} + n_\mathrm{T}$ being lower at the plasma core in {\it LoPed} than in {\it HiPed}.
The mix of deuterium and tritium is optimal in both scenario variants, i.e., $n_\mathrm{D} = n_\mathrm{T}$.

For convenience, differences in the safety factor $q$ will be discussed later, apropos figure \ref{figure_most_unstable_modes_in_gap} and figure \ref{figure_q_detail}.

\subsection{Numerical codes and modelling workflow}
\label{section_modelling}

Our workhorse is \mbox{CASTOR-K}, which is used to assess the stability of AEs through the computation of their linear growth-rates.
Besides \mbox{CASTOR-K} our suite of numerical codes comprises HELENA \cite{Huysmans1991} to obtain magnetic equilibria, and the incompressible ideal-MHD code MISHKA \cite{Mikhailovskii1997} to compute the eigenmodes of a given equilibrium.
The extensive identification of all the AEs of an equilibrium is key to the success of our methodical stability-analysis approach.
For this reason MISHKA has been preferred to the resistive-MHD code CASTOR, which we also use to calculate radiative-damping rates, as reported in section \ref{section_radiative_damping}.
In a simple performance test done by executing both MHD codes in the same CPU, with the same input, MISHKA calculated an eigenmode in around 2.5 s, while CASTOR took approximaely 23.5 s to compute the same eigenmode.
MISHKA is therefore roughly 10 times faster than CASTOR, which makes it an adequate tool to solve for thousands of eigenmodes in a reasonably short time.

We focus on eigenmodes with toroidal mode number $n$ ranging from 1 to 50 in order to stay within the limits of the drift-kinetic ordering for alpha particles, i.e., $k_\perp\rho_\alpha<1$ where $k_\perp$ is the AE perpedicular wavenumber and $\rho_\alpha$ is the gyroradius of the alpha particles \cite{Rodrigues2015}.
We further restrict our analysis to AEs whose eigenfrequency falls in one of the first three gaps of the ideal shear Alfv\'en wave continuum, namely (ordered by increasing frequency) the Toroidicity induced AE (TAE) gap, the Ellipticity induced AE (EAE) gap, and the Non-circular triangularity-induced AE (NAE) gap \cite{Fu1989, Betti1991, Kramer1998}.
All possible TAEs, EAEs, and NAEs with $n\leq 50$ have been determined by scanning the mode frequency $\omega/\omega_\mathrm{A}$ from $0.01$ to $2.0$, a value higher than the top frequency of the NAE gap for the range of $n$ considered.
The scan step is $2.0×\times10^{-5}$, which we have found to be sufficiently small not to miss eigenmodes.
At every step of the scan, the quantity $(\omega/\omega_\mathrm{A})^2$ is input to MISHKA as a guess of the mode eigenvalue which, upon convergence, is returned together with the corresponding mode eigenfunction.
Valid AEs are subsequently collected from the large set of eigenmodes produced by the frequency scan. 
The selection is based on two criteria, namely, the eigenfunction must be well-resolved radially \cite{Rodrigues2015}, and the mode cannot be affected by continuum damping. 
We account for the effect of continuum damping in a straightforward, binary way.
A given AE is considered to be fully damped if its frequency matches the ideal Alfv\'en continuum at any radial position where the mode amplitude exceeds 1\% of its maximum, otherwise continuum damping is not taken into consideration at all for that mode.
A total of 705 AEs have successfully passed this validation process for {\it LoPed}, and 401 for {\it HiPed}.
The selected AEs are then processed by \mbox{CASTOR-K}, which calculates the energy $\delta W_\mathrm{p}$ exchanged between a mode and a given population p of plasma particles.
The associated contribution to the growth rate, $\gamma_\mathrm{p}=\mathrm{Im}(\delta W_\mathrm{p})/(2\omega W_\mathrm{k})$, where $W_\mathrm{k}$ is the kinetic energy of the mode perturbation \cite{Borba1999} is computed by \mbox{CASTOR-K} as well --- it is the basic quantity used in the stability analysis of eigenmodes.

Four \mbox{CASTOR-K} runs have been done for each mode to calculate the drive due to the alpha particles ($\alpha$) and the Landau damping due to the interaction with the bulk ions (DT), the electrons (e) and the helium ashes (He).
The net growth rate of the AE is then obtained by summing these 4 contributions, i.e., 
\begin{equation}
\gamma = \gamma_\alpha + \gamma_\mathrm{DT} + \gamma_\mathrm{e} + \gamma_\mathrm{He}. 
\label{equation_net_growth_rate}
\end{equation}
Note that for some modes, particularly for even, Low-Shear TAEs (LSTAEs) \cite{Fu1995} which sit near the bottom of the TAE gap and consist of a symmetric combination of poloidal harmonics (see figure \ref{figure_LSTAE_sequences}) \cite{Nyqvist2012}, radiative damping may have an additional non-negligible contribution to $\gamma$, a subject that will be addressed in section \ref{section_radiative_damping}.

In \mbox{CASTOR-K}, the distribution function of every particle population p is modelled with the product of a function of $s$ and a function of energy $E$ \cite{Nabais2015, Rodrigues2015},
\begin{equation}
F_\mathrm{p}(s,E)=n_\mathrm{p}(s)f_\mathrm{p}(E).
\label{equation_distribution_function}
\end{equation}
The radial profiles $n_\mathrm{p}(s)$ of the thermal populations (DT, e and He), and of the alpha particles are shown in figure \ref{figure_scenario} for {\it LoPed} and {\it HiPed}.
Concerning the energy distribution $f_\mathrm{p}(E)$, while DT ions, He ashes and electrons have been described by Maxwellian distributions \cite{Rodrigues2015}, a slowing-down distribution which is determined by the effects of electron and ion drag on alpha particles \cite{Gaffey1976, Candy1996, Pinches1998, Bilato2014} has been used to describe the alpha-particle population,
\begin{equation}
f_\mathrm{\alpha}(E)=f_\mathrm{sd}(E)\left/\int_0^\infty f_\mathrm{sd}(E)dE\right., 
\label{equation_normalized_slowing_down}
\end{equation}
where
\begin{equation}
f_\mathrm{sd}(E)=\frac{1}{E^{\,3/2}+E_c^{\,3/2}}\,\mathrm{erfc}\left(\frac{E-E_0}{\Delta E}\right).
\label{equation_slowing_down}
\end{equation}
This expression provides a good approximation to distributions calculated with Fokker-Planck models \cite{Gaffey1976, Bilato2014}, and its analytical simplicity is convenient for the calculation of derivatives in Castor-K.
The crossover energy $E_\mathrm{c}$ is the alpha-particle energy below which ion drag becomes more important than electron drag.
It is given by
\begin{equation}
E_\mathrm{c}=T_\mathrm{e}\left(\frac{3Z_1}{4}\right)^{2/3}\left(\frac{\pi m_\alpha}{m_\mathrm{e}}\right)^{1/3},
\label{equation_crossover_energy}
\end{equation}
where $Z_1=\sum_i m_\mathrm{\alpha}n_i z_i^2/\left(m_in_\mathrm{e}\right)$ is a sum over ions, the $i$th ion species having density $n_i$, charge number $z_i$ and mass $m_i$, and the electron temperature $T_\mathrm{e}$ is measured in eV.
Using values of $T_\mathrm{e}$ at $s=0.4$ where the gradient of $n_\alpha$ is practically at its maximum we obtain $E_\mathrm{c} = 595.1$ keV in {\it LoPed} and $E_\mathrm{c} = 423.2$ keV in {\it HiPed}.
The ion temperature $T_\mathrm{i}$ at $s=0.4$ has been chosen as the dispersion of the birth energy of alpha particles around the value $E_0=3.5$ MeV, i.e., $\Delta E = 15.5$ keV in {\it LoPed} and $\Delta E = 11.4$ keV in {\it HiPed}.

Sensitivity analysis showed that varying $\Delta E$ from 10 keV to 100 keV changed $\left|\gamma_\alpha\right|$ by at most 0.5\% in {\it LoPed} and 0.1\% in {\it HiPed}.
Concerning the sensitivity to variations of $E_\mathrm{c}$, using $T_\mathrm{e}$ values in equation \ref{equation_crossover_energy} taken in the region of strong $n_\alpha$ gradient $0.25\lesssim s\lesssim 0.55$ led to a maximum variation in $\left|\gamma_\alpha\right|$ of 10\% in {\it LoPed} and 5\% in {\it HiPed}.

\section{Modelling results}
\label{section_results}

This section is essentially focused on the characterization of the destabilized eigenmodes with particular emphasis on their growth rates, which are reported in section \ref{section_growth_rates}.
Since the evaluation of the radiative-damping contribution to the net growth-rates is much less automated or systematic than the calculation of the other drive and damping terms,
it is discussed separately in section \ref{section_radiative_damping}. 
The radial structure and frequency distribution of AEs within the TAE gap is discussed in section \ref{section_TAE_gap}.

\subsection{AE stability}
\label{section_growth_rates}

\begin{figure*}
\begin{center}
\includegraphics[width=1.0\textwidth]{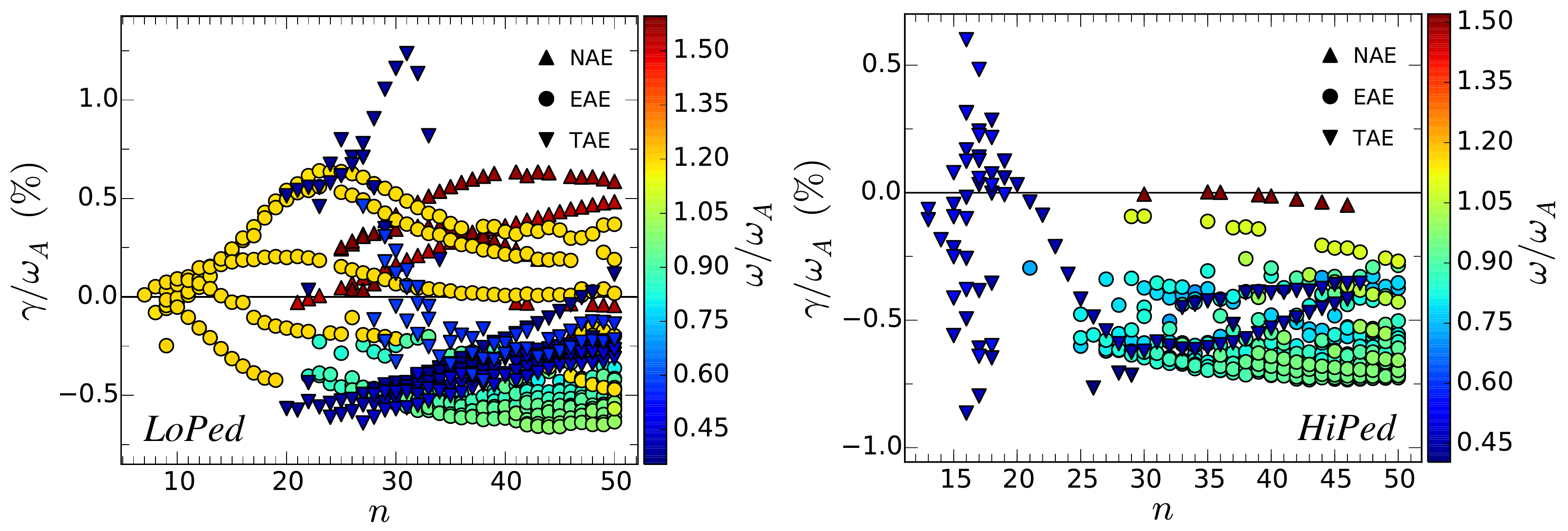}
\caption{Linear growth rates $\gamma$ normalized to the Alfv\'en frequency $\omega_\mathrm{A} = v_\mathrm{A}/R_\mathrm{m}$, where the Alfv\'en velocity at the magnetic axis is $v_\mathrm{A} \approx 7.1\times10^6$ m/s in {\it LoPed} and $v_\mathrm{A} \approx 7.0\times 10^6$ m/s in {\it HiPed}, versus toroidal mode number $n$ and colored by AE frequency for the two variants of the 15 MA ITER baseline scenario. TAEs appear in dark-blue patterns which correspond to frequencies around $\omega_\mathrm{A}/2$ and are the most unstable.}
\label{figure_growth_rates_all_AE}
\end{center}
\end{figure*}
The linear growth rates computed with \mbox{CASTOR-K} for all the valid AEs found in both variants of the ITER baseline scenario are represented in figure \ref{figure_growth_rates_all_AE}, where different symbols are used for TAEs, EAEs, and NAEs (notice the normalization to the Alfv\'en frequency instead of the more common mode frequency).
A quite noticeable feature of the growth rates in figure \ref{figure_growth_rates_all_AE} is that they are larger in {\it LoPed} than in {\it HiPed}, which is consistent with the much higher alpha-particle density and density gradient in the {\it LoPed} variant of the scenario --- see figure \ref{figure_scenario}.
Moreover, it is striking that although a large number of EAEs and NAEs have positive growth rates in {\it LoPed}, clearly in both scenario variants all markedly unstable modes are TAEs.
No AEs have been found for $n<7$ in {\it LoPed} and $n<13$ in {\it HiPed}.

\begin{figure*}
\begin{center}
\includegraphics[width=1.0\textwidth]{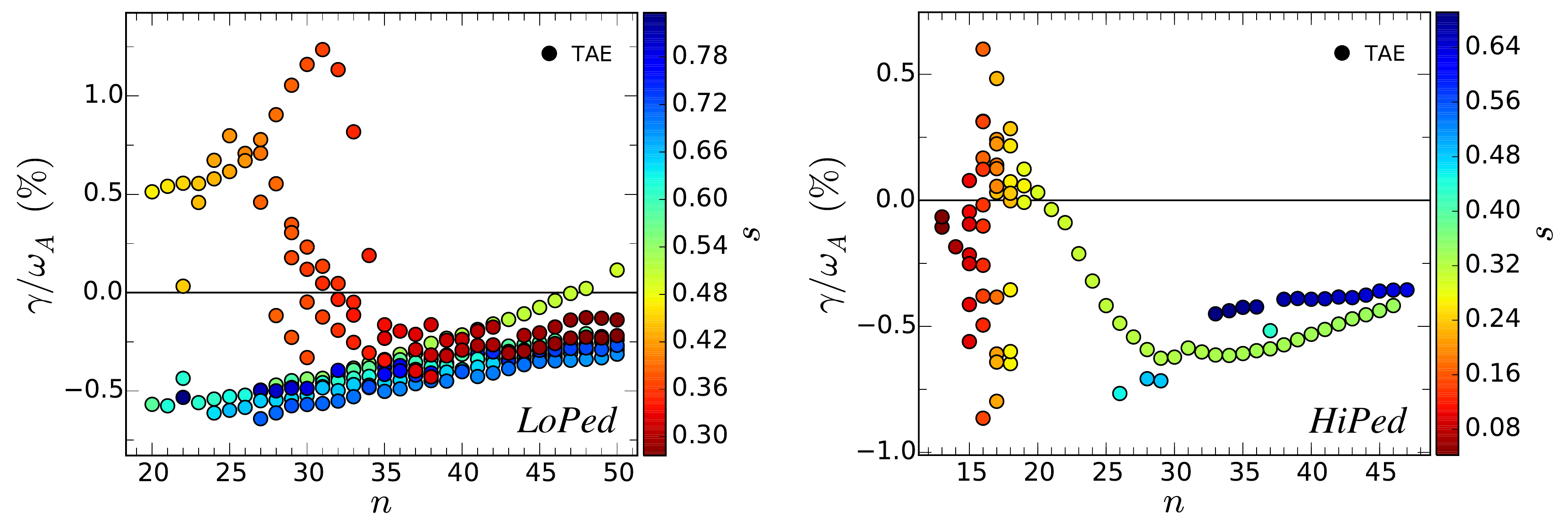}
\caption{In both scenarios analysed the most unstable TAEs are core-localized, as indicated by their reddish colors. As seen in the $q$ profile in figure \ref{figure_most_unstable_modes_in_gap}, in {\it HiPed} these TAEs are inside the $q=1$ surface (i.e., they are ``tornado modes'' \cite{Sharapov2013}).}
\label{figure_growth_rates_TAE}
\end{center}
\end{figure*}

\begin{figure*}
\begin{center}
\includegraphics[width=1.0\textwidth]{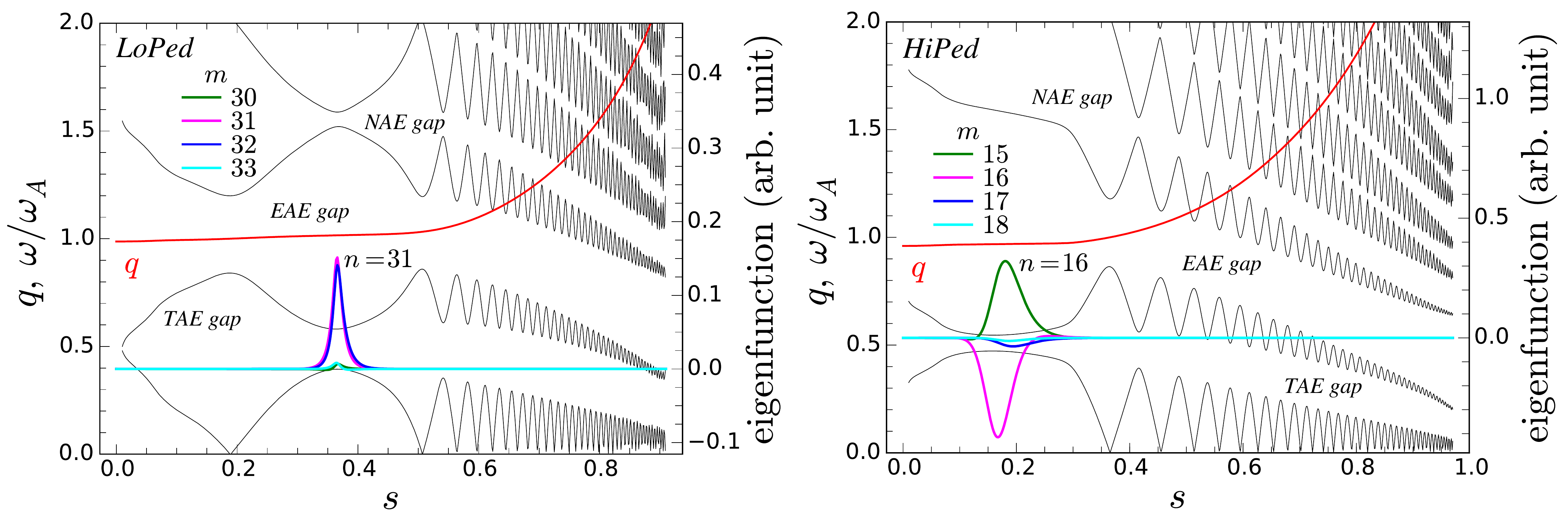}
\caption{The eigenfunctions of the most unstable AEs represented by the colored lines have been computed by MISHKA using 17 harmonics with poloidal mode number starting at $m=n-1$, but here only the strongest 4 are shown. In both scenario variants the modes fall within the TAE gap of the Alfv\'en continuous spectrum, whose boundaries are shown as solid black lines --- notice that the baseline of the eigenfunction (zero value on the right axis) marks the mode frequency on the left axis.
}
\label{figure_most_unstable_modes_in_gap}
\end{center}
\end{figure*}

\begin{figure}
\begin{center}
\includegraphics[width=0.49\textwidth]{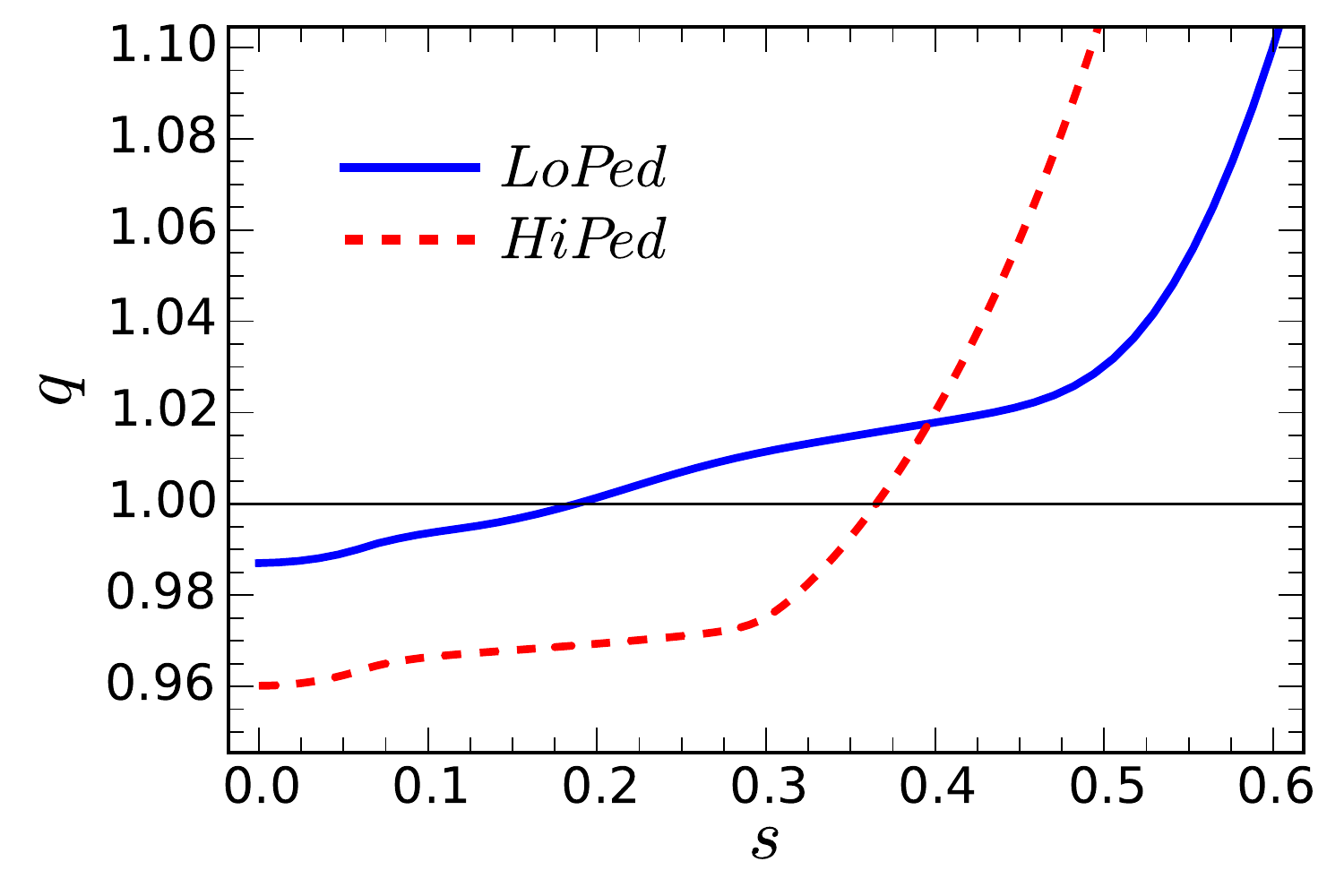}
\caption{The on-axis value of the safety factor in {\it HiPed} is $q(0) = 0.96$, around 3\% lower than $q(0) = 0.987$ in {\it LoPed}. While in {\it HiPed} the low magnetic-shear region (due to sawtooth crashes) ranges from the axis to $s \approx 0.3$, in {\it LoPed} it extends to $s \approx 0.5$. The $q=1$ surface is located at $s \approx 0.2$ in {\it LoPed} and $s \approx 0.37$ in {\it HiPed}.
}
\label{figure_q_detail}
\end{center}
\end{figure}

Figure \ref{figure_growth_rates_TAE} shows the TAEs from figure \ref{figure_growth_rates_all_AE} with their radial location, which is defined as the position of the maximum amplitude of their strongest poloidal harmonic.
It is clear that the unstable modes are well localized at the core of the plasma, inside the region $s\lesssim 0.48$ for {\it LoPed} and $s\lesssim 0.32$ for {\it HiPed}.
It can further be seen from the radial profile of $q$ in figure \ref{figure_most_unstable_modes_in_gap} that these modes exist within the low magnetic-shear region of the plasma --- they are in fact LSTAEs.
Only a few modes exist for $n\gtrsim 20$ in {\it HiPed}.
These are low-frequency modes sitting close to the bottom of the TAE gap which do not cross the continuous spectrum, as seen in figure \ref{figure_TAE_gap_distribution}.
Such is not the case in {\it LoPed} for which more modes exist for high $n$.
This is due to the extended low magnetic-shear region, within which LSTAEs with higher eigenfrequencies can exist without matching the top boundary of the TAE gap, as seen in figure \ref{figure_most_unstable_modes_in_gap}, and because in {\it LoPed} higher-$n$ modes are located at inner positions than lower-$n$ modes.
Indeed, a difference in the evolution of the location of LSTAEs as $n$ increases can also be seen in figure \ref{figure_growth_rates_TAE}.
While in {\it HiPed} modes become progressively located at positions farther from the core, the opposite occurs in {\it LoPed}.
This contrast in behavior results from $q(s)$ being monotonically increasing (see figure \ref{figure_most_unstable_modes_in_gap}) together with the fact that within the $0.2<s<0.37$ region $q$ is below 1 in {\it HiPed} but above 1 in {\it LoPed}, as shown in figure \ref{figure_q_detail}.
Considering the TAE condition $q(s)=(m\pm 1/2)/n$ \cite{Cheng1986} and that $m\approx n$ for LSTAEs we obtain $q(s)\approx 1\pm 1/(2n)$.
Therefore, since $q>1$ in {\it LoPed}, $q$ must decrease as $n$ increases so TAEs move towards lower $s$ values, whereas for $q<1$ in {\it HiPed} TAEs move towards higher $s$ values as $q$ increases.
Moreover, the higher magnetic shear in {\it HiPed} does not allow the existence of LSTAEs beyond $s \approx 0.3$, that is for $n\gtrsim20$.

The growth rates of the unstable TAEs in figure \ref{figure_growth_rates_TAE} are visibly peaked at toroidal mode numbers that are roughly $n \approx 30$ in {\it LoPed} and $n \approx 15$ in {\it HiPed}.
As shown in figure \ref{figure_most_unstable_modes_in_gap} the most unstable mode in {\it LoPed} is a $n = 31$ even LSTAE radially localized at $s\approx 0.37$, whose frequency $\omega/\omega_\mathrm{A}=0.395$ lies at $0.5\%$ of the TAE gap ($0\%$ being the bottom gap frequency and $100\%$ the top gap frequency). 
This mode has a net growth rate $\gamma/\omega_\mathrm{A} \approx 1.24\%$ of which $2.17\%$ is from alpha-particle drive and $-0.88\%$ is due to damping by the bulk ions. 
In {\it HiPed} the most unstable mode is a $n = 16$ LSTAE found at $s\approx 0.17$ with a frequency $\omega/\omega_\mathrm{A}=0.533$, which lies at $83.1\%$ of the TAE gap.
It is an odd mode, formed by an anti-symmetric combination of poloidal harmonics (see figure \ref{figure_LSTAE_sequences}) \cite{Nyqvist2012}, and its growth rate is $\gamma/\omega_\mathrm{A} \approx 0.60\%$, for which alpha-particles contribute with $0.94\%$ and bulk ions with $-0.33\%$.
As happens in these two examples, Landau damping of TAEs is in general substantial (as discussed in detail in section \ref{section_TAE_gap}) and it is mainly due to the DT ions.
Indeed, we have found damping by He ashes to be relatively small and damping by electrons to be negligible.
It should be noticed that while in {\it LoPed} the most unstable mode is close to $s=0.38$ where the gradient of the alpha-particle density $n_\alpha$ is highest, that is not possible in {\it HiPed} because the maximum $n_\alpha$ gradient occurs at $s\approx 0.44$ where the higher magnetic shear only allows non-local modes that interact strongly with the Alfv\'en continuum. 
Nevertheless, in both scenario variants the $n_\alpha$ gradient remains close to its maximum value in the mid-radius region $0.25\lesssim s\lesssim 0.55$ which encloses all unstable AEs.

At this point it is interesting to verify a well-known estimate for the toroidal mode number of the most driven AE \cite{Pinches2015, Rodrigues2015}.
The estimate is based on matching the width of passing alpha-particle orbits and the TAE width, which leads to $n\approx s/q^2\times a/R_\mathrm{m} \times \Omega_{\alpha}/\omega_\mathrm{A}$, where the cyclotron frequency of the alpha particles is $\Omega_{\alpha} \approx 2.5\times10^8$ rad/s and $q(s)$ is evaluated at the location $s$ of the AE with the highest drive.
By using $q(0.37) \approx 1.016$ we arrive at $n \approx 26$ in {\it LoPed}, while $q(0.17) \approx 0.968$ in {\it HiPed} leads to $n \approx 13$. 
Considering the simplicity of the rationale behind them, these are reasonable guesses for the toroidal mode numbers of the most driven, and for that matter also of the most unstable AEs. 
However, it must be remarked that the above formula depends significantly on the location of the most driven modes, which could not have been known with accuracy beforehand.

\subsection{Radiative damping}
\label{section_radiative_damping}

With the introduction of finite gyroradius effects TAEs become coupled with Kinetic Alfv\'en Waves (KAWs), which carry energy in the radial direction away from the eigenmode in an intrinsic TAE damping process known as radiative damping \cite{Rosenbluth1975, Mett1992, Candy1994, Nyqvist2012, Pinches2015}.
In the case of LSTAEs only the symmetric modes suffer significant radiative damping, as the KAWs excited by the anti-symmetric LSTAEs interfere within the mode localization region resulting in a negligible net energy flux \cite{Mett1992, Nyqvist2012}.

Radiative damping has previously been shown to have a significant contribution to the net growth rate $\gamma$ in {\it LoPed} \cite{Rodrigues2015}, namely for its most unstable mode, a symmetric LSTAE sitting very close to the bottom of the TAE gap.
On the contrary, the most unstable modes in {\it HiPed} are anti-symmetric LSTAEs, so their radiative damping is expected to be very small \cite{Nyqvist2012}.
Here, the same approach that has been followed in \cite{Rodrigues2015} is used to estimate $\gamma_\mathrm{rad}$, the radiative-damping contribution to the net growth rate of selected AEs from both scenario variants.
The method \cite{Gorelenkov1992, Mett1992, Candy1994, Connor1994} relies on a formal equivalence between a non-ideal MHD model that accounts for finite parallel electric field and first-order ion-gyroradius effects, and the resistive MHD model that is implemented in the CASTOR eigenvalue code \cite{Kerner1998}. 

In order to compute non-ideal eigenmodes and determine their radiative-damping rates, in place of the (usually real) resistivity we input to CASTOR the complex quantity \cite{Candy1994, Connor1994}
\begin{equation}
\eta = i 4 q^2 \left\{ \frac{3}{4} + \frac{T_\mathrm{e}}{T_\mathrm{i}} \left[ 1-i\delta \left(\nu_\mathrm{e}\right) \right] \right\} \left( \frac{\omega}{\omega_\mathrm{A}} \right)^3 \left(  \frac{\rho_\mathrm{i}}{R_\mathrm{m}} \right)^2,
\label{equation_complex_resistivity}
\end{equation}
and conduct a scan in $\delta(\nu_\mathrm{e})$, a wave dissipation-rate due to collisional friction between trapped electrons and passing particles ($\nu_\mathrm{e}$ is the electron collision frequency), which leads to an imaginary frequency component and to the corresponding damping rate $\gamma_\mathrm{CASTOR}$.
We thus obtain the MHD growth-rate $\gamma_\mathrm{CASTOR}$ as a function of $\delta(\nu_\mathrm{e})$, from which $\gamma_\mathrm{rad}$ can be inferred.

Our method is based on the fact that of the two components of non-ideal eigenmodes, KAWs suffer much stronger collisional damping than AEs \cite{Hasegawa1976, Candy1994}.
Therefore, as $\delta(\nu_\mathrm{e})$ is increased from zero $\gamma_\mathrm{CASTOR}$ rises mainly due to the dominant damping of KAWs, up to a certain value of $\delta(\nu_\mathrm{e})$ for which all KAW energy is damped by collisional friction and only the much weaker damping of the AE remains.
This change of behavior is observed as a modification of the slope of the $\gamma_\mathrm{CASTOR}$ versus $\delta(\nu_\mathrm{e})$ curve which shows a noticeable knee, as discussed below.
An indication of the scan limits can be taken from known expressions of $\delta(\nu_\mathrm{e})$, for which $0<\delta(\nu_\mathrm{e})\ll 1$ \cite{Candy1994, Connor1994}.
It should therefore not be necessary to extend the scan to $\delta(\nu_\mathrm{e})>1$ to observe the knee.

\begin{figure*}
\begin{center}
\includegraphics[width=1.0\textwidth]{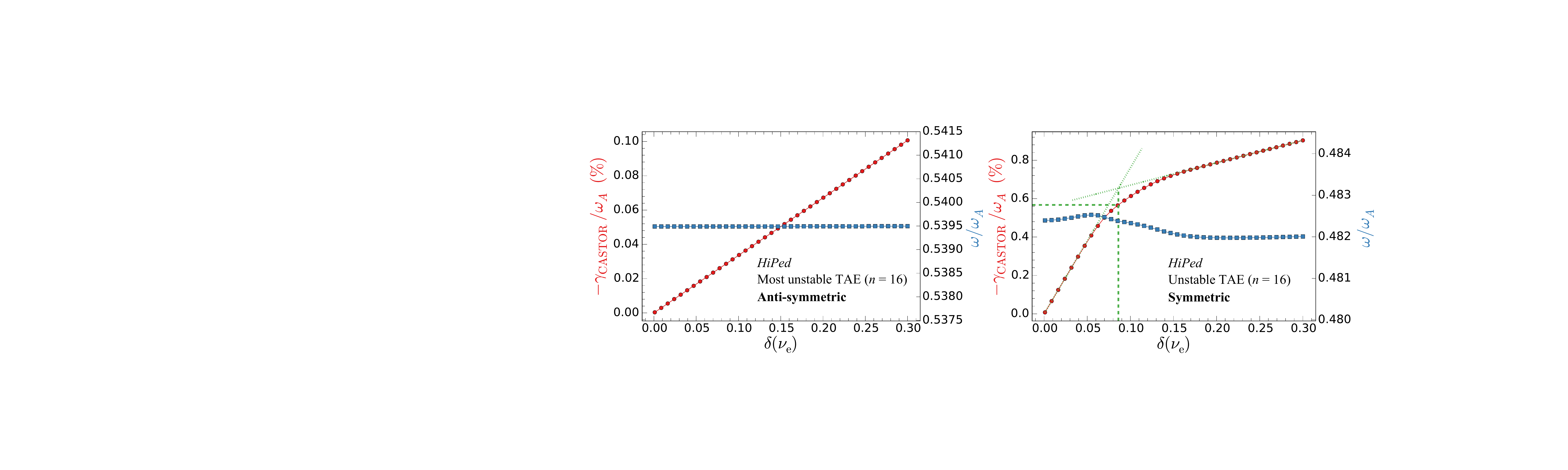}
\caption{Estimation of radiative-damping rates for a pair of {\it HiPed} TAEs. Red curves represent the MHD damping-rate calculated by CASTOR as a function of the wave dissipation-rate $\delta(\nu_\mathrm{e})$ \cite{Candy1994, Connor1994}, whereas blue curves represent the frequency of the non-ideal eigenmode. For the analysis of the TAE on the left, which is the most unstable in {\it HiPed}, the values $T_\mathrm{e}=14.7\,\mathrm{keV}$ and $T_\mathrm{i}=13.0\,\mathrm{keV}$ are used in equation \ref{equation_complex_resistivity}. The constant slope of the red curve indicates a negligible value of $\gamma_\mathrm{rad}$, as expected since this is an anti-symmetric mode \cite{Nyqvist2012}. In the case of the symmetric TAE on the right the temperatures are $T_\mathrm{e}=14.8\,\mathrm{keV}$ and $T_\mathrm{i}=13.1\,\mathrm{keV}$. The 5-point BIS analysis of the red curve leads to a radiative-damping rate $\gamma_\mathrm{rad}/\omega_\mathrm{A} \approx -0.57\%$, which has been picked at $\delta_\mathrm{rad}\approx0.09$ as indicated by the green lines.} 
\label{figure_radiative_damping}
\end{center}
\end{figure*}

The right-hand side of equation \ref{equation_complex_resistivity} is evaluated at the position of the eigenmode in question.
At the location of all the eigenmodes of interest the normalized ion gyroradius $\rho_\mathrm{i}/R_\mathrm{m}$ has the value $6.25×\times10^{-4}$ in {\it LoPed} and $3.43×\times10^{-4}$ in {\it HiPed}, whereas $q$ is approximately $1.02$ in {\it LoPed} and $0.97$ in {\it HiPed}.
For the AE frequency $\omega$ in equation \ref{equation_complex_resistivity} we use the value given by MISHKA.
Notice that this frequency would only be the same as the frequency of the eigenmode given by CASTOR if $\eta=0$, i.e., in the ideal-MHD case if compressibility is negligible.
In our calculations we found that the CASTOR frequency is generally slightly higher, by up to 5\%, than the MISHKA frequency.
Moreover, as can be seen in figure \ref{figure_radiative_damping} the frequency of the CASTOR eigenmode changes during a $\delta(\nu_\mathrm{e})$ scan --- except when radiative damping is very small.
To address these frequency changes and ensure that we are indeed analysing the intended AE, a scan is made not only in $\delta(\nu_\mathrm{e})$ but also in the guess frequency that is input to CASTOR.
To initiate a $\delta(\nu_\mathrm{e})$ scan we use a guess frequency that is close to the MISHKA eigenmode frequency. Subsequently, the guess frequency at every scan step is given by the frequency of the converged CASTOR eigenmode in the previous step of the scan.
This way it is guaranteed that the input given to CASTOR changes slowly and we are tracking the same eigenmode during the whole $\delta(\nu_\mathrm{e})$ scan, which considerably simplifies the process.
By scanning a range of initial guess-frequency values around the frequency of the MISHKA eigenmode we obtain a set of $\delta(\nu_\mathrm{e})$ scans.
From this set we can then choose the scan for the non-ideal CASTOR eigenmode that corresponds to the desired ideal MISHKA eigenmode with coupled KAWs.

The $\delta(\nu_\mathrm{e})$ scan on the left of figure \ref{figure_radiative_damping} is for the most unstable TAE in {\it HiPed}, which we recall is anti-symmetric.
The constant slope of $\gamma_\mathrm{CASTOR}/\omega_\mathrm{A}$ versus $\delta(\nu_\mathrm{e})$ exemplifies the outcome of a $\delta(\nu_\mathrm{e})$ scan for odd LSTAEs, which do not suffer noticeable radiative damping.
As discussed above, this is an expected result given the negligible energy carried away by the KAWs, and it has been observed for the other anti-symmetric and noticeably unstable modes in {\it HiPed} as well.
Loosely speaking, in this case the MHD damping rate is solely due to the weak damping of the ideal AE component of the non-ideal eigenmode, and it is therefore simply proportional to the wave dissipation-rate $\delta(\nu_\mathrm{e})$.
The eigenfunctions calculated by CASTOR and by MISHKA are practically the same for these eigenmodes, as well as their frequencies which differ by no more than 1\%.

The right side of figure \ref{figure_radiative_damping} shows the $\delta(\nu_\mathrm{e})$ scan for the most unstable {\it symmetric} TAE in {\it HiPed} with the same $n=16$. 
This eigenmode has a net growth-rate $\gamma/\omega_\mathrm{A} \approx 0.31\%$.
It is localized at $s \approx 0.15$ and its frequency is $\omega/\omega_\mathrm{A}=0.481$.
In contrast with the TAE analysed on the left side of figure \ref{figure_radiative_damping}, the CASTOR eigenmode frequency $\omega/\omega_\mathrm{A}$ now varies noticeably during the scan.
Simultaneously, as $\delta \left(\nu_\mathrm{e}\right)$ rises there is a significant change in the slope of the $\gamma_\mathrm{CASTOR}/\omega_\mathrm{A}$ versus $\delta(\nu_\mathrm{e})$ curve, which eventually reaches a constant value as the curve asymptotically approaches a straight line.
A break-in-slope (BIS) technique is used to pinpoint $\delta_\mathrm{rad}$, the value of $\delta(\nu_\mathrm{e})$ at the intersection of the two lines obtained by linear fitting the first and last few points of the curve, as shown by the green lines in figure \ref{figure_radiative_damping}.
Following the reasoning above, the ordinate at the knee of the curve (where its slope changes abruptly), which in this case is for the abscissa $\delta_\mathrm{rad}\approx0.09$ is taken as the radiative-damping rate $\gamma_\mathrm{rad}/\omega_\mathrm{A} \approx -0.57\%$.
Although the net growth rate calculated by \mbox{CASTOR-K} for this particular AE is positive, the eigenmode is in fact stable when considering radiative damping, since $(\gamma+\gamma_\mathrm{rad})/\omega_\mathrm{A} \approx -0.26\%$.
Notice that these are somewhat rough estimates of the radiative-damping rate $\gamma_\mathrm{rad}$ that depend on several factors.
In particular they depend on the values of $T_\mathrm{e}$ and $T_\mathrm{i}$ used in equation \ref{equation_complex_resistivity}, which are calculated at a particular value of $s$ chosen to represent the radial position of the eigenmode.
The uncertainty in $\delta_\mathrm{rad}$ must also be considered.
Evidently, a different method could be used to determine the knee of the $\gamma_\mathrm{CASTOR}/\omega_\mathrm{A}$ versus $\delta(\nu_\mathrm{e})$ curve.
We could for example choose the point where the slope of $\gamma_\mathrm{CASTOR}/\omega_\mathrm{A}$ and the frequency $\omega/\omega_\mathrm{A}$ become practically stable, $\delta(\nu_\mathrm{e})\approx 0.15$. 
The radiative-damping rate would in that case differ from the BIS value by approximately 20\%.
Nevertheless, such a difference would not have a significant impact on the stability of the TAEs.

Table \ref{table_radiative_damping} summarizes the radiative-damping analysis that has been made for some of the most unstable eigenmodes in {\it LoPed}, all of which are symmetric.
These modes have been chosen from the top of the two curves that can be seen on the left of figure \ref{figure_growth_rates_TAE} peaking around $n=25$ and $n=31$. 
While the analysis confirms that radiative damping accounts for a significant fraction of the net growth-rate in {\it LoPed}, it also shows that the most unstable eigenmode remains the same after taking radiative damping into account, as other unstable eigenmodes suffer a similar radiative-damping effect. 
\begin{table*}
\begin{center}
\caption{\label{table_radiative_damping}Radiative-damping contribution to the net growth-rate for a selection of unstable LSTAEs in {\it LoPed}.}
\captionsetup{justification=centering}
\begin{tabular}{@{}*{9}{c}@{}}
\hline
\noalign{\smallskip}
$n$ & $\omega/\omega_\mathrm{A}$ & $s$ & $T_\mathrm{e}\,\mathrm{(keV)}$ & $T_\mathrm{i}\,\mathrm{(keV)}$ & $\gamma/\omega_\mathrm{A}$ & $\delta_\mathrm{rad}$ & $\gamma_\mathrm{rad}/\omega_\mathrm{A}$ & $(\gamma+\gamma_\mathrm{rad})/\omega_\mathrm{A}$ \\
\noalign{\smallskip}
\hline 
\noalign{\smallskip}
24 & $0.382$ & $0.43$ & $17.0$ & $14.8$ & $0.67\%$ & $0.12$ & $-0.63\%$ & $0.04\%$ \\
25 & $0.384$ & $0.42$ & $17.3$ & $15.0$ & $0.80\%$ & $0.12$ & $-0.50\%$ & $0.30\%$ \\
26 & $0.386$ & $0.41$ & $17.6$ & $15.3$ & $0.71\%$ & $0.12$ & $-0.71\%$ & $0.00\%$ \\
30 & $0.394$ & $0.37$ & $18.7$ & $16.2$ & $1.16\%$ & $0.14$ & $-0.73\%$ & $0.43\%$ \\
31 & $0.395$ & $0.37$ & $18.7$ & $16.2$ & $1.24\%$ & $0.09$ & $-0.36\%$ & $0.88\%$ \\
32 & $0.397$ & $0.36$ & $19.0$ & $16.4$ & $1.13\%$ & $0.07$ & $-0.41\%$ & $0.72\%$ \\
\noalign{\smallskip}
\hline
\end{tabular}
\end{center}
\end{table*}

\subsection{Radial structure and frequency distribution of LSTAEs}
\label{section_TAE_gap}

\begin{figure}
\begin{center}
\includegraphics[width=0.49\textwidth]{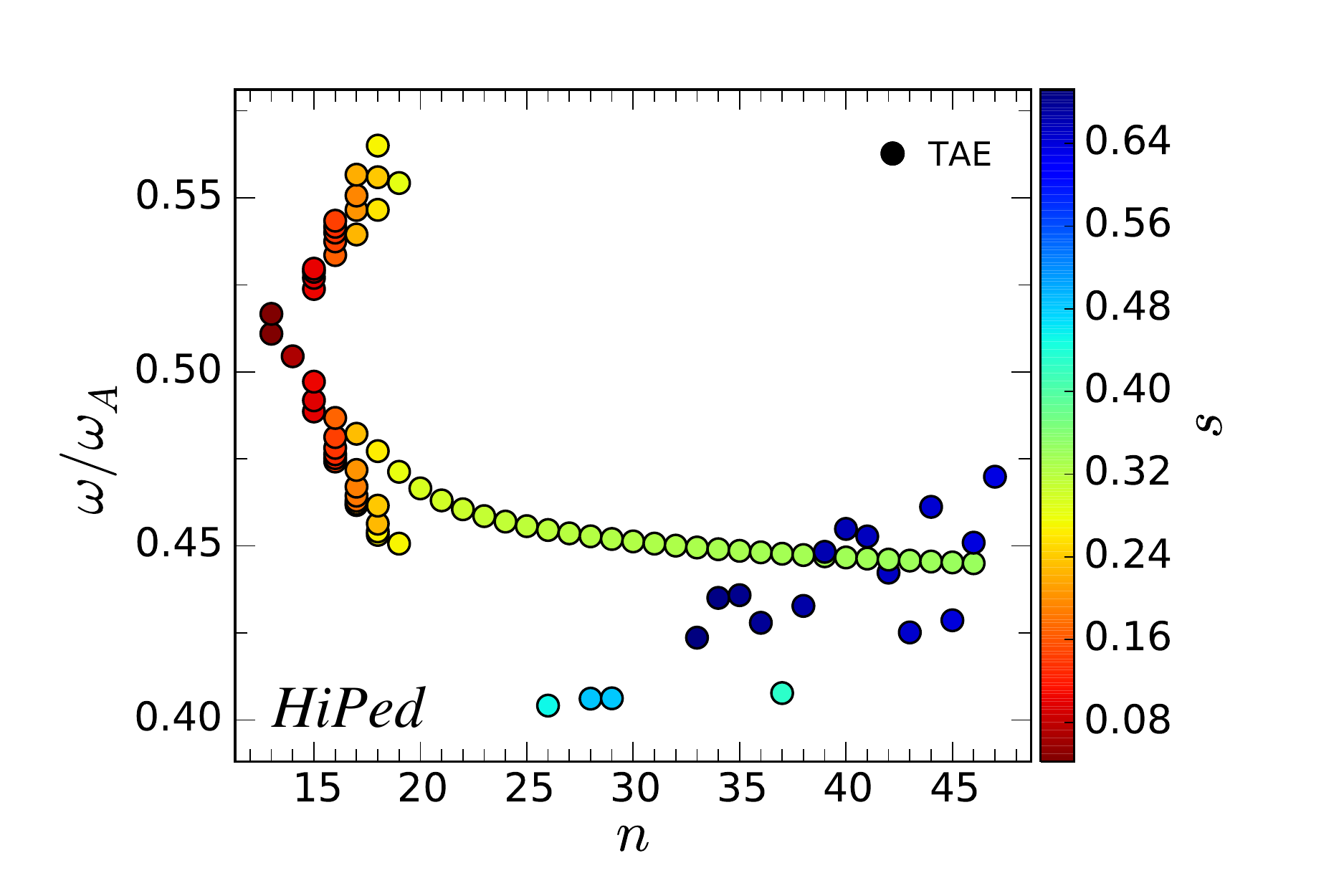}
\caption{Frequency distribution of LSTAEs as a function of toroidal mode number in {\it HiPed}. The lower branch is formed by symmetric modes near the bottom of the TAE gap, while the modes in the upper branch are anti-symmetric and close to the top of the gap. The modes that appear scattered at the bottom right of the figure with $s\gtrsim 0.4$ are not LSTAEs.}
\label{figure_TAE_gap_distribution}
\end{center}
\end{figure}
In figure \ref{figure_TAE_gap_distribution} the frequency of TAEs is plotted versus $n$ for {\it HiPed}.
Two main branches are evident inside the TAE gap: an upper branch made of modes that rise in frequency as $n$ increases, and a lower branch with decreasing frequency modes.
The lower-frequency branch is made of even, symmetric modes, while the modes in the upper branch are odd, anti-symmetric modes \cite{Pinches2015, Lauber2015}.
Furthermore, it has been verified that for a given $n$ the frequency of anti-symmetric modes rises with the number of peaks in their poloidal harmonics, while symmetric modes have progressively lower frequencies as their number of peaks increases.
\begin{figure*}
\begin{center}
\includegraphics[width=1.0\textwidth]{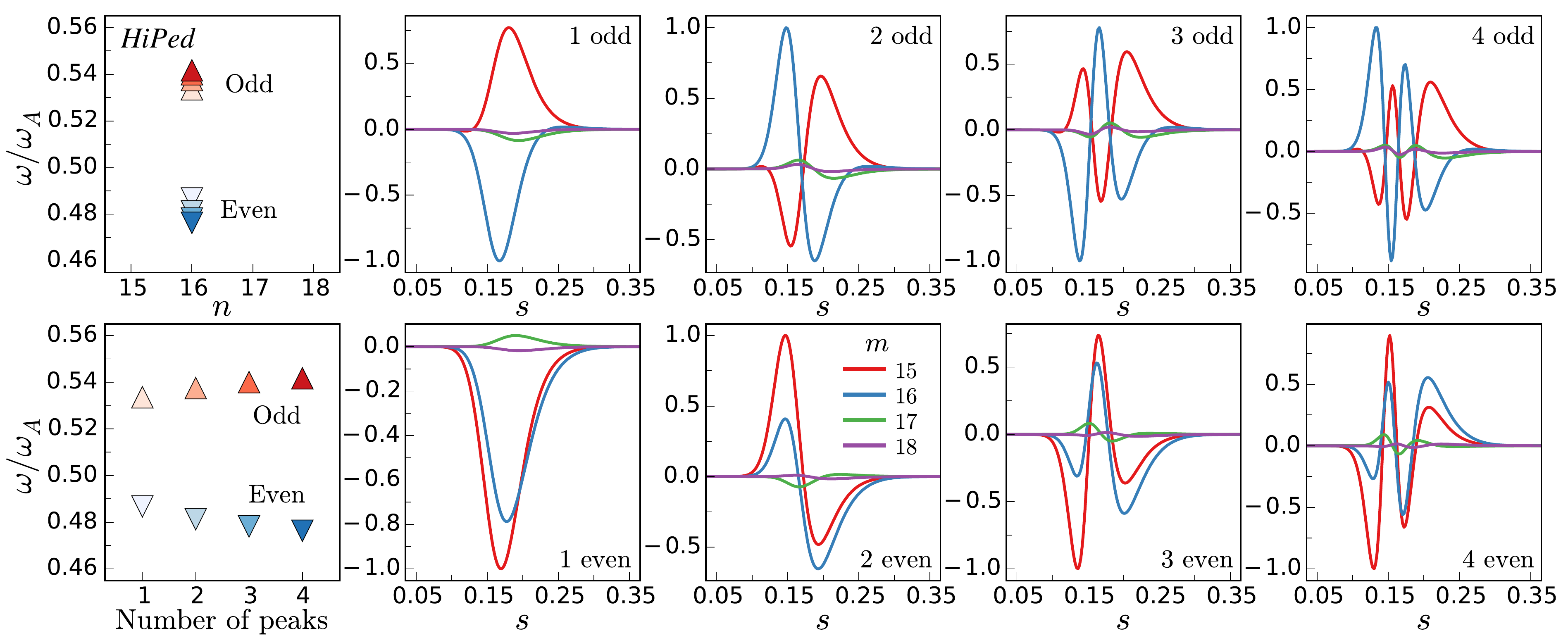}
\caption{This group of $n=16$ LSTAEs in {\it HiPed} is a subset of the data in figure \ref{figure_TAE_gap_distribution}. The sequence in the top row shows the eigenfunctions of odd, anti-symmetric modes, whereas the eigenfunctions in the bottom row form a sequence of even, symmetric modes. The absolute value of every eigenfunction has been normalized to unity. Moreover, eigenfunctions have been labelled with the number of peaks in each of their poloidal harmonics, of which only the first 4 are shown, and with their parity.}
\label{figure_LSTAE_sequences}
\end{center}
\end{figure*}
This behavior is illustrated in figure \ref{figure_LSTAE_sequences} for $n=16$, but it occurs for all the modes in the two branches of figure \ref{figure_TAE_gap_distribution} and it has also been observed in {\it LoPed}.

The long line in figure \ref{figure_TAE_gap_distribution} that chirps down until $n \approx 45$ has the simplest symmetric TAEs with a single peak, no zeroes or oscillations.
Its ``mirror'' line in the upper branch ends abruptly at $n = 19$.
This occurs because, as discussed earlier, in {\it HiPed} higher-$n$ modes are located farther from the core than lower-$n$ modes and their eigenfrequencies match the Alfv\'en continuum at a smaller $s$ for modes at the top of the gap than at its bottom, as can be seen in figure \ref{figure_most_unstable_modes_in_gap}, thereby causing the missing anti-symmetric modes to be effectively annihilated by continuum damping. 

\begin{figure*}
\begin{center}
\includegraphics[width=1.0\textwidth]{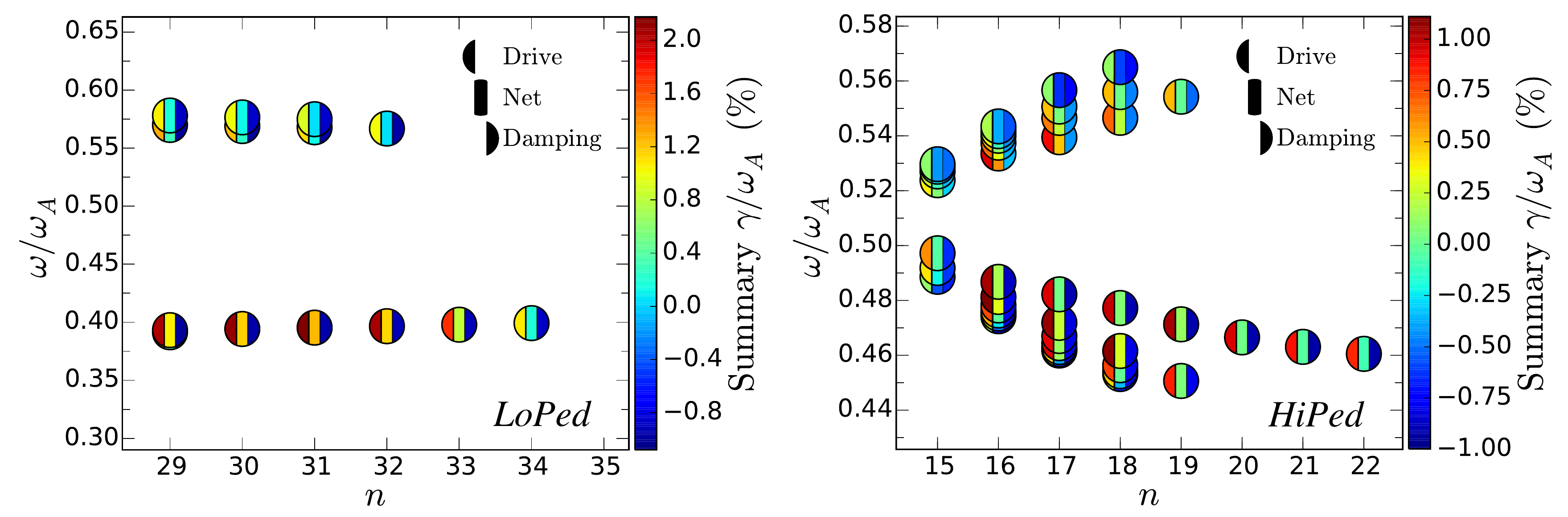}
\caption{Summary of growth-rates for symmetric and anti-symmetric LSTAEs in both scenario variants, showing their net value $\gamma$ (middle part of the circular markers), drive $\gamma_\alpha$ (left marker part), and damping $\gamma_\mathrm{DT} + \gamma_\mathrm{e} + \gamma_\mathrm{He}$ (right marker part) --- see equation \ref{equation_net_growth_rate}. Only modes with positive net growth-rates are shown. In {\it LoPed} the most driven modes are the symmetric LSTAEs, which are also the most unstable. That is not the case in {\it HiPed} because although symmetric modes are the most driven, they also have the highest damping, whereby the most unstable {\it HiPed} modes are anti-symmetric LSTAEs.}
\label{figure_symmetry}
\end{center}
\end{figure*}

The opposite situation occurs in {\it LoPed} as higher-$n$ modes are located at positions closer to the core than lower-$n$ modes, so the anti-symmetric modes are missing for $n < 27$.
Figure \ref{figure_most_unstable_modes_in_gap} also shows that in both scenario variants the most unstable modes are LSTAEs with a single peak. 
However, whereas in {\it LoPed} symmetric modes have larger growth rates than anti-symmetric modes with the same $n$, in {\it HiPed} Alfv\'enic activity is dominated by anti-symmetric modes.
This situation is unusual since the TAEs that are commonly observed in experiments are symmetric \cite{Kramer2004}. 
The leading role of anti-symmetric TAEs in {\it HiPed} is explained in figure \ref{figure_symmetry}, where it can be seen that it is due to their lower damping and not to a lower alpha-particle drive of the symmetric modes.

\section{Summary}
\label{section_conclusions}

The linear stability of AEs in the presence of alpha particles has been analysed for two variants of the 15 MA ELMy H-mode ITER baseline scenario using a specialized workflow that is based on the hybrid MHD drift-kinetic code \mbox{CASTOR-K}.
Our modelling results show that, considering alpha-particle drive and Landau damping on DT ions, helium ashes and electrons, the most unstable modes have toroidal mode numbers around $n \approx 30$ in the scenario variant with a lower pedestal temperature, named {\it LoPed}, and $n \approx 15$ in {\it HiPed}, with maximum growth rates of 1.24\% and 0.60\%, respectively.
In both scenario variants these modes are LSTAEs, i.e., they are localized in the low magnetic-shear region of the plasma core.
{\it LoPed} has a higher density of alpha particles at the plasma core than {\it HiPed}, which is consistent with the larger AE growth rates that have been found for {\it LoPed}.

Radiative damping, which has been determined for a number of chosen modes in both scenarios, was shown to somewhat reduce the growth rates of the most unstable modes in {\it LoPed}, but to be insufficient to stabilize them or alter their growth-rate ordering.
This result is in line with the calculation done in \cite{Rodrigues2015} for the most unstable mode in {\it LoPed}.
On the contrary, {\it HiPed} results are essentially unaltered since in this case all significantly unstable modes are anti-symmetric TAEs, which are practically unaffected by radiative damping \cite{Nyqvist2012}. 
Nevertheless, radiative damping having been considered, Alfv\'enic activity remains most unstable in the {\it LoPed} variant of the ITER baseline scenario. 

Concerning symmetry, in the case of {\it HiPed} a clear frequency distribution of symmetric and anti-symmetric LSTAEs within the TAE gap has been found that agrees with and illustrates results from recent studies on the same ITER scenario \cite{Pinches2015, Lauber2015}.
It has been found that practically all unstable modes are symmetric in {\it LoPed}, whereas in {\it HiPed} anti-symmetric modes have the highest growth rates.
The rather uncommon fact that in {\it HiPed} the most unstable modes are not symmetric has been shown to be due to the lower Landau-damping rates of the anti-symmetric modes.
Indeed, the most driven modes in {\it HiPed} are symmetric LSTAEs just like in {\it LoPed}, which shows the importance of considering all drive and damping processes when assessing the stability of a scenario.

Neutral beam injection (NBI), which has not been considered here, has previously been found to drive AEs quite significantly for $s\gtrsim 0.3$ in the case of {\it LoPed}, assuming a 1 MeV birth energy of the NBI fast ions \cite{Pinches2015, Toigo2015}.
It is therefore important to calculate the NBI drive, certainly so in the case of {\it LoPed} for which the most unstable AEs are located at $s\gtrsim 0.3$.
This shortcoming is to be addressed in future work. 

It is important to note that differences in the safety factor, namely in the radial extent of the low magnetic-shear region have been shown to strongly influence both the number of existing AEs and their radial location, particularly of the most unstable LSTAEs.
Moreover, since magnetic shear is very low in most of the plasma core, some care must be taken in the interpretation of these and similar stability results as they can be sensitive to relatively small variations in the on-axis value of the safety factor \cite{Rodrigues2015EPS, Rodrigues2015IAEA}.

\section*{Acknowledgments}

This work has been carried out within the framework of the EUROfusion Consortium and has received funding from the Euratom research and training programme 2014-2018 under grant agreement No. 633053.
IST activities received financial support from ``Funda\c{c}\~{a}o para a Ci\^{e}ncia e Tecnologia'' (FCT) through project UID/FIS/50010/2013.
ITER is Nuclear Facility INB no. 174.
The views and opinions expressed herein do not necessarily reflect those of the European Commission, IST, FCT, or the ITER Organization.
All computations were carried out using the HELIOS supercomputer system at the Computational Simulation Centre of the International Fusion Energy Research Centre (IFERC-CSC) in Aomori, Japan, under the Broader Approach collaboration between Euratom and Japan implemented by Fusion for Energy and JAEA.
PR was supported by EUROfusion Consortium grant no. WP14-FRF-IST/Rodrigues and NFL was supported by FCT grant no. IF/00530/2013.

\providecommand{\newblock}{}

\end{document}